\RequirePackage{fix-cm}
\documentclass[smallextended]{svjour3}       
\smartqed  
\usepackage{graphicx}
\usepackage{epstopdf}
\usepackage{hyperref}
\usepackage{subfig}
\usepackage{comment}
\usepackage{mathtools}
\usepackage[labelfont=bf,labelsep=space]{caption}
\usepackage{enumitem}
\usepackage{float}
\begin{document}

\title{Automatic Weighted Matching Rectifying Rule Discovery for Data Repairing
}


\subtitle{Can we discover effective repairing rules automatically from dirty data?}


\author{Hiba Abu Ahmad   \and
        Hongzhi Wang
}


\institute{Hiba Abu Ahmad \at
              1. Department of Computer Science, Harbin Institute of Technology, Harbin, China\\
              2. Department of Software Engineering and Information Systems, Tishreen University, Latakia, Syria\\
              \email{hiba.abu.ahmad@hit.edu.cn}
           \and
            Hongzhi Wang \at
              Department of Computer Science, Harbin Institute of Technology, Harbin, China\\
              \email{wangzh@hit.edu.cn}
}


\date{Received: date / Accepted: date}

\maketitle

\begin{abstract}
Data repairing is a key problem in data cleaning which aims to uncover and rectify data errors. Traditional methods depend on data dependencies to check the existence of errors in data, but they fail to rectify the errors. To overcome this limitation, recent methods define repairing rules on which they depend to detect and fix errors. However, all existing data repairing rules are provided by experts which is an expensive task in time and effort. Besides, rule-based data repairing methods need an external verified data source or user verifications; otherwise they are incomplete where they can repair only a small number of errors. In this paper, we define weighted matching rectifying rules (WMRRs) based on similarity matching to capture more errors. We propose a novel algorithm to discover WMRRs automatically from dirty data in-hand. We also develop an automatic algorithm for rules inconsistency resolution. Additionally, based on WMRRs, we propose an automatic data repairing algorithm (WMRR-DR) which uncovers a large number of errors and rectifies them dependably. We experimentally verify our method on both real-life and synthetic data. The experimental results prove that our method can discover effective WMRRs from the dirty data in-hand, and perform dependable and full-automatic repairing based on the discovered WMRRs, with higher accuracy than the existing dependable methods.

\end{abstract}

\section{Introduction}
\label{section:introduction}
Data quality is one of the most crucial problems in data management. Database systems usually concentrate on the data size aiming to create, maintain, and control a great volume of data. However, real-life data is often dirty and poor quality; about 30\% of firms' data could be dirty \cite{fullofdirtydata}. Dirty data is very expensive where its expenses exceed 3 trillion dollars for the USA economy \cite{dirtydata}. In addition, high-quality data are so critical in decision making. These indicate and emphasize the necessity of data cleaning for organizations. Data repairing is a key problem in data cleaning to uncover data errors and rectify these errors.

Different data dependencies are proposed for data repairing, such as functional dependencies (FDs)\cite{bohannon2005cost}, conditional functional dependencies (CFDs)\cite{bohannon2007conditional}, matching dependencies (MDs)\cite{fan2009reasoning}, and lately conditional matching dependencies (CMDs)\cite{wang2017discovering}. Although data dependencies can judge if errors exist in the data or not, they fail to determine wrong values, and worse, they cannot fix the wrong values. For that, various types of rules are defined to detect and fix errors, which are editing rules \cite{fan2010towards,fan2012towards}, fixing rules \cite{wang2014towards} and Sherlock rules \cite{interlandi2015proof}. Even though the rules-based data repairing methods \cite{fan2010towards,fan2012towards,interlandi2015proof} outperform data dependencies-based methods, they need an exterior trustworthy data source or users verification. 
In contrast, fixing rules-based method \cite{wang2014towards} performs data repairing without using master data or involving users, but it can repair only a small number of data errors. Furthermore, all proposed rules for data repairing are provided by domain experts, which is a long-time, impractical and costly task.

We explain the limitations of existing methods in Example \ref{example:motivated example}.

\begin{example}\label{example:motivated example} Consider the data set $D_{RES}$ in Table \ref{table:motivated example table} of researchers including 8 tuples: $t_1,t_2,\dots,t_8$ where each tuple $t_i$ refers to a researcher, identified by $Name$, $Department (Dept)$, $Nationality (Nation)$ and $Capital$. The symbol ``$*$'' signs all errors whose corrections are given between brackets, for example, $t_2(Capital)$= ``HongKong'' is an error, whose correction is ``Beijing''.

Suppose a functional dependency $fd: RES$ $(Nation \rightarrow Capital)$ over $D_{RES}$, which indicates that $Nation$ uniquely specifies $Capital$. Since $(t_1,t_2)$ violates $fd$ where $t_1(Nation)=t_2(Nation)$ but $t_1(Capital)\neq t_2(Capital)$. Thus, $fd$ confirms that it must be errors in the values: $t_1(Nation)$, $t_2(Nation)$, $t_1(Capital)$, $t_2(Capital)$, but it cannot determine which values are wrong or how they can be fixed.

Consider a tuple s in a master data $D_m$ as follows:

\begin{table}[!ht]
  \centering
    \begin{tabular}{lll}
     \hline\noalign{\smallskip}
    $TupleID$ & $Country$ & $Capital$ \\
    \noalign{\smallskip}\hline\noalign{\smallskip}
    $s$     & China & Beijing \\
    \hline\noalign{\smallskip}
    \end{tabular}
    \label{table:master data}
\end{table}

On both data sets ($D_{RES}, D_m$), experts can define an editing rule $er$: $((Nation, Country) \rightarrow (Capital,Capital),t_p=())$. It denotes that: for an input tuple $t \in D_{RES}$ in Table \ref{table:motivated example table}, if $t(Nation)$ is correct (a user verifies it) and $\exists s \in D_m$ where  $t(Nation)= s(Country)$ and $t(Capital) \neq s(Capital)$; $t(Capital)$ is wrong which is fixed to $s(Capital)$. Accordingly, $er$ detects errors $t_2(Capital)$ and $t_4(Capital)$ and updates them to ``Beijing''. 
Moreover, since $er$ depends on exact matching, it cannot fix $t_6(Capital)$, $t_3(Nation)$, and $t_6(Nation)$. 

\begin{table}
  \caption{$D_{RES}$: An instance of the schema $RES$ }
    \begin{tabular}{lllll}
      \hline\noalign{\smallskip}
    $TupleID$&$Name$  & $Dept$  & $Nation$ & $Capital$ \\
    \noalign{\smallskip}\hline\noalign{\smallskip}
   $t_1$ & Wu    & CS    & China & Beijing \\
   $t_2$ & Li    & CS    & China & HongKong* (Beijing) \\
   $t_3$ & Kum   & AI    & Chiena*(China) & Beijing \\
   $t_4$ & Shi   & AI    & China & Shanghai* (Beijing) \\
   $t_5$ & Xu    & MC    & China & Beijing \\
   $t_6$ & Pei   & MC    & Chiena*(China) & HongKong* (Beijing) \\
   $t_7$ & Wei   & CS    & China & Beijing \\
   $t_8$ & Wang   & CS    & China & Beijing \\
      \hline\noalign{\smallskip}
    \end{tabular}
    \label{table:motivated example table}
\end{table}

Similarly, consider a Sherlock rule $sr$: $((Nation,Country),(Capital,\bot, $ $Capital) \rightarrow (=,\not\approx,=))$, defined on ($D_{RES}, D_m$). It depends on the similarity instead of the equality to compare $t(Nation)$ with $s(Country)$, $\forall t\in D_{RES}, s\in D_m$. Therefore, for Table \ref{table:motivated example table}, it can annotate each correct value as positive, each wrong value as negative, and update wrong values to correct ones. However, Sherlock rules, as editing rules, are provided by experts using master data. 

In opposite, consider a fixing rule, which is not based on master data to define or users verification to apply, $fr$: $((Nation, Capital),(Capital,$                           \{``Hongkong'', ``Shanghai''\}) $\rightarrow$ ``Beijing''), also provided 
by experts. $fr$ requires an evidence on $Nation$ attribute with correct value, e.g., ``China'' to fix wrong values in $Capital$, e.g., ``Hongkong'' and ''Shanghai'' updating them to ``Beijing''. Therefore, it can fix $t_2(Capital)$ and $t_4(Capital)$, but it fails to detect or fix $t_6(Capital)$, $t_3(Nation)$ and $t_6(Nation)$. $\qed$
\end{example}
Note that fixing rules, which are the only kind of rules for automated and dependable data repairing, require valid evidence from some attributes to detect and fix errors in other related attributes. So, if there is even a typo in the evidence, the errors in the related attributes, as well as that typo, cannot be detected and fixed.

This paper introduces weighted matching rectifying rules to overcome the previous limitations. Example \ref{example:Matching rectifying rules} discusses the cases that these rules can cover.

\begin{example}
\label{example:Matching rectifying rules}
Consider the data set $D_{RES}$ in Table \ref{table:motivated example table}, $Nation$ and $Capital$ as two related attributes based on $fd$. We notice, first, that for a particular value of $Nation$, e.g., ``China'', the correct value of $Capital$, i.e. ``Beijing'' has more frequency than the wrong ones, e.g., ``Hongkong'' and ``Shanghai''. Second, using approximately valid values of a set of attributes can help us to detect and fix more errors in the related attributes. Based on these two notices, we generate a weighted matching rectifying rule $r$: $((Nation \approx$ ``China''$),(Capital \in $ \{``Hongkong'',``Shanghai''\}) $\Rightarrow ($ ``China''$ ) \land ($ ``Beijing''$))$. This rule can detect and rectify errors not only in the related attribute, $Capital$, but also in the evidence attribute, $Nation$, as follows:

\begin{itemize}
\item $t_2(Nation)$= ``China'' and $t_2(Capital) \in$ \{``Hongkong'',``Shanghai''\}, then $t_2(Capital)$ is wrong and we update it to ``Beijing''. Similarly, $t_4(Capital)$ is rectified.
\item $t_6(Nation) \approx$ ``China'' and $t_6(Capital) \in$ \{``Hongkong'',``Shanghai''\}, then $t_6(Capital)$ is wrong and we update it to ``Beijing'', and $t_6(Nation)$ is wrong and we update it to ``China''.
\item $t_3(Nation) \approx$ ``China''  and $t_3(Capital)$= ``Beijing'', then $t_3(Nation)$ is wrong and we update it to ``China''. $\qed$
\end{itemize}

\end{example}

The two examples above raise the following challenges to develop a matching rectifying rules-based data repairing method:
\begin{itemize}
\item how to define weighted matching rectifying rules over the dirty data in-hand with more flexible matching as they can detect and fix different errors dependably and automatically?
\item how to discover these rules automatically from the dirty data in-hand?
\item What is the effective method to apply these rules for automated and dependable data repairing?
\end{itemize}

Consider these challenges, the main contributions of this paper are summarized as follows:

\begin{itemize}
\item We define weighted matching rectifying rules (WMRRs) that can cover and fix more data errors dependably and automatically.
\item We propose an automatic rule discovery algorithm based on the dirty data in-hand and their functional dependencies. According to our knowledge, it is the first automatic rule discovery method for data repairing.
\item We study fundamental problems of WMRRs, and develop an automatic algorithm to check rules consistency and resolve rules inconsistency.
\item We propose an effective data repairing method based on a consistent set of WMRRs.
\item We conduct comprehensive experiments on two data sets, which verify the effectiveness of our proposed method.
\end{itemize}

The rest of this paper is organized as follows. Section \ref{section:realted work} studies related work. Section \ref{section:Weighted Matching Rectifying Rules} defines WMRRs. Section \ref{section:Rule Discovery} presents the automatic rule discovery algorithm. Section \ref{section:Fundamental Problems} studies the fundamental problems of WMRRs. Section \ref{section:Rule Inconsistency Resolution} demonstrates the automatic algorithm for rules inconsistency resolution. Section \ref{section:WMRR-based Data Repairing} presents the automatic repairing algorithm based on WMRRs. Section \ref{section:Experiment Results} reports our experimental results, and finally, the paper is concluded in Section \ref{section:conclusion}.







\section{Related Work}
\label{section:realted work}
Many studies have addressed data cleaning problems, especially data repairing which can be classified as follows.

\emph{Dependencies-based Data Repairing.} Heuristic data repairing based on data dependencies have been broadly proposed \cite{bohannon2005cost,cong2007improving,fan2008dependencies}. They addressed the problem of exploring a consistent database with a minimum difference from the original database \cite{arenas1999consistent}. They used various cost functions to fix errors and employed different dependencies such as FDs \cite{kolahi2009approximating} \cite{beskales2010sampling}, CFDs \cite{beskales2014sampling,fan2008conditional}, CFDs and MDs \cite{fan2014interaction}, and DCs \cite{chu2013holistic}. However, consistent data is not necessarily correct. Therefore, the proposed solutions by these methods can not guarantee correctness.

\emph{Rule-based Data Repairing.} Unlike dependencies-based methods, rule-based methods are more dependable and conservative. Therefore, rules have been developed for different data cleaning problems such as ER-rules for entity resolution \cite{ahmad2018effective,li2014rule}, editing rules\cite{fan2012towards}, fixing rules \cite{wang2014towards} and Sherlock rules \cite{interlandi2015proof} for data repairing. Fixing rules can be discovered by users interaction \cite{he2016interactive}, or provided by experts \cite{wang2014towards} like editing rules \cite{fan2012towards} and Shelock rules \cite{interlandi2015proof}. In contrast, our proposed rules, WMRRs, are discovered automatically based on the data in-hand. From another side, editing rules \cite{fan2012towards} depend on master data and user verifications to perform reliable repairing, while Sherlock rules depend on master data for automatic repairing. Related to this study, fixing rules \cite{wang2014towards} perform reliable and automatic repairing without external data sources. However, the repairing process based on fixing rules is incomplete, i.e., only a little number of errors can be fixed, where fixing rules focus on repairing correctness at the expense of repairing completeness. In opposite, weighted matching rectifying rules focus on both completeness and correctness of repairing. Moreover, WMRR-data repairing is reliable and automatic. We will explain experimentally (Sect. \ref{section:Experiment Results}) how WMRR-based data repairing can significantly improve the recall with maintaining the precision of repairing.

\emph{Data Repairing using Knowledge Bases.} Some methods have utilized knowledge bases for data repairing. KATARA \cite{chu2015katara} used knowledge bases and crowdsourcing to detect correct and wrong values, so it is a nonautomatic method. For rule discovery, \cite{Hao2018} used knowledge bases to generate deductive rules (DRs) which can identify correct and wrong values, and fix errors only if there is enough evidence. However, rule discovery needs enough correct and wrong record examples to investigate the right and error semantics of the data from the knowledge base, and the expensive expert knowledge is still necessary to validate the extracted semantics. For data repairing, \cite{Hao2018} requires to design effective semantic links between dirty databases and knowledge bases which is user-guided, i.e., nonautomatic. In contrast, we aim automatic rule discovery and automatic data repairing based on the data in-hand utilizing correct data to fix wrong data without any external source.

\emph{User Guided Data Repairing.} Since users and experts can help to perform reliable repairing, they were involved in various data repairing method \cite{raman2001potter,heer2015predictive,yakout2011guided}, even in rule-based methods \cite{he2016interactive,interlandi2015proof,Hao2018} as we discussed before. However, depending on users is commonly costly in terms of effort and time, and worse error-prone, while domain experts are not always available with the required knowledge. Accordingly, automatic data repairing is needed which we target in this work.

\emph{Machine Learning and statistical-based Data Repairing.} Machine Learning are also employed by some data repairing methods, such as \cite{rekatsinas2017holoclean,yakout2013don}. These methods are particulary supervised since they require training data and rely on the chosen features. Other methods perform statistical repairing by applying probabilistic to infer the correct data \cite{shin2015incremental,bach2015hinge,niu2011tuffy}. Thus, our method, as a rule-based method, varies from this class of methods in that it is a declarative method to determine correct values and repair wrong values automatically based on the data in-hand.

Indeed, rule-based data cleaning methods are often preferred by end users, because rules are explicable, simply rectify and refine \cite{singh2017generating}. As a result, they have been diffused in industries and business, such as ETL (Extract-Transform-Load) rules \cite{herzog2007data}. However, there is still an essential need for automatic rule discovery and automatic methods based on the rules which this work introduces.

\section{Weighted Matching Rectifying Rules}
\label{section:Weighted Matching Rectifying Rules}
In this section, we introduce weighted matching rectifying rules for data repairing, WMRRs. Consider a data set $D$ over a schema $S$ with a set of attributes $A=\{a_1, a_2,\dots,a_n\}$, where each attribute $a_i$ has a finite domain $dom(a_i)$. We first define the syntax of the rules. Then, we describe the semantics of the rules.

\subsection{Rule Syntax}
\label{subSection:Rule Syntax}
A matching rectifying rule $r$ defined on a schema $S$ has the following syntax:

$[X \approx_X DP(X)]\land [y \in WP(y)]\Rightarrow [DP(X)] \land [cp(y)]$, where

\begin{itemize}
\item $X \subset A$ is a set of attributes in schema $S$, and $y \in A \setminus X$ is an attribute in $R$ but not in $X$;
\item $DP(X)$ is a pattern with attributes X, called as the \emph{director pattern} on X such that $\forall a \in X, DP(a) \in dom(a)$ is a constant value in the domain of attribute $a$;
\item $WP(y)\subset dom(y)$ is a vector of constant values in the domain of attribute $y$, called as the \emph{wrong patterns} of $y$;
\item $cp(y) \in dom(y)\setminus WP(y)$ is a constant value in the domain of y but not in $WP(y)$, called as the \emph{correct pattern} of $y$.
\item $\approx_X$ is a similarity metric on attributes $X$ that identifies the similarity between $X$ values in a rule, i.e., $DP(X)$ and the corresponding values of $X$ in a tuple, i.e. $t(X)$. $DP(X) \approx_{X} t(X)$ iff $DP(x_i) \approx_{x_i} t(x_i)$ $\forall  x_i \in X$. Formally, $\approx_{x_i}$ indicates true or false as follows.

\begin{equation}
\label{Eq.Similarity metric}
  DP(x_i) \approx_{x_i} t(x_i)=\begin{cases}
    true, & \text{if $sim(DP(x_i),t(x_i))<\vartheta $}\\
    false, & \text{otherwise},
  \end{cases}
\end{equation}
\end{itemize}

where  $sim(DP(x_i),t(x_i))$ is a similarity function, and $\vartheta $ is a threshold.

\emph{Similarity Function.} $\approx_X$ can use domain-specific similarity operators or any similarity functions, like Edit distance, Jaccard distance, Cosine similarity and Euclidean distance, with a predefined threshold $\vartheta$. To check similarity, by default, Edit distance is used for attributes with string values and Euclidean distance for attributes with numeric values \cite{hao2017novel}. In this paper, we formally have
\begin{equation}
\label{Eq.Similarity function}
  sim(DP(x_i),t(x_i)) =\begin{cases}
    editD(DP(x_i),t(x_i)), & \text{string}.\\
    euclD(DP(x_i),t(x_i)), & \text{numeric}.
  \end{cases}
\end{equation}


\emph{Rule Weights.} Since we discover rules from dirty data, rules could be, in turn, dirty. We assign two weights for each rule $r$: $w_1(r)$ and $w_1(r)$, to assure a good performance of the rules on $D$ for data repairing. $w_1(r)$, which is used for rule discovery (Sect. \ref{section:Rule Discovery}) and rule inconsistency resolution (Sect. \ref{section:Rule Inconsistency Resolution}), measures the validity of the rule. $w_2(r)$, which is used for rule-based data repairing (Sect. \ref{section:WMRR-based Data Repairing}), measures the ratio of tuples with correct values for both attributes $X$ and $y$ to all tuples in the data set $D$. We define the weights of a rule $r$ as follows:
\begin{equation}
\label{weight1 definition}
w_1(r)=\frac{|DP(X)\cup cp(y)|_D}{|DP(X)|_D},
\end{equation}

\begin{equation}
\label{weight2 definition}
w_2(r)=\frac{|DP(X)\cup cp(y)|_D}{|D|},
\end{equation}
where $|DP(X)\cup cp(y)|_D$ denotes the number of tuples in $D$ with $DP(X)$ and $cp(y)$ values for the attributes $X$ and $y$, respectively, $|DP(X)|_D$ denotes the number of tuples in $D$ with $DP(X)$ values for the attributes $X$, and $|D|$ denotes the data $D$ size in terms of tuples.

$w_1(r) \in$ [0,1] is (a) the probability that $y$ attribute has a wrong value in a tuple $t \in D$ and will be rectified to $cp(y)$, or $X$ set of attributes has one attribute or more with a typo in a tuple $t \in D$ which will be rectified to $CP(X)$ when $t$ matches $r$; or (b) the probability that a tuple $t \in D$ has a correct value $cp(y)$ for $y$ attribute, and a correct value $CP(X)$ for $X$ set of attributes when $t$ matches $r$.

$w_2(r) \in$ [0,1] is the probability that correct values $CP(X)$ and $cp(y)$ of attributes $X$ and $y$, respectively, appears together in data set tuples.

For example, $w_1(r)=2/3$  and $w_2(r)=1/2$ are the weights of the rule $r$ in Example \ref{example:Matching rectifying rules} based on Eq. (\ref{weight1 definition}) and Eq. (\ref{weight2 definition}), respectively.


\subsection{Rule Semantics}
\label{subSection:Rule Semantics}
Let $t$ be a tuple in a data set $D$, and $r \in R$ be a weighted matching rectifying rule with the syntax in Sect. \ref{subSection:Rule Syntax}. Intuitively, $X$ and $y$ are semantically correlated. $R$ is a consistent set of weighted matching rectifying rules. The following definitions describe the semantics of applying the rule $r$, and the rule set $R$.

\begin{definition}
$t$ matches $r$, denoted by $t\vdash r$, if (1) $t(X)\approx_X DP(X)$, and $t(y)\in WP(y)$, or (2) $t(X)\approx_X DP(X)$, and $t(y)=cp(y)$.
\end{definition}

\begin{definition}
$r$ is applied to $t$ if $t$ matches $r$, changing $t$ to $\acute{t}$, denoted by $t \rightarrow_{r} \acute{t}$, where $\acute{t}(x)=CP(x)$ $\forall x \in X$ and $\acute{t}(y)=cp(y)$. This includes: (1) $r$ rectifies $X$ if $\exists x \in X; t(x)\neq cp(x)$, (2) $r$ rectifies $y$ if $t(y)\in WP(y)$, (3) $r$ verifies $x \in X$ if $t(x)=CP(x)$, then $\acute{t}(x)=t(x)$, and (4) $r$ verifies $y$ if $t(y)=cp(y)$, then $\acute{t}(y)=t(y)$.
\end{definition}

Therefore, $r$ can rectify wrong values and verify correct values of $t(X)$ and $t(y)$ when $t$ matches $r$.

\begin{example}
Consider the data set in Table \ref{table:motivated example table} and the rule $r$ in Example \ref{example:Matching rectifying rules}. $t_2,t_4,$ and $t_6$ match $r$ since $t_i(Nation)\approx$ ``China'', and $t_i(Capital) \in$ \{``Hongkong'', ``Shanghai''\} $\forall i \in \{2,4,6\}$. $t_3$ also matches $r$ since $t_3(Nation) \approx$ ``China'', and $t_3(y)$= ``Beijing''. Consequently, $r$ detects and fixes all errors in these tuples as follows: $t_2(Capital),t_4(Capital), t_6(Capital)$ are updated to ``Beijing'', and $t_3(Nation)$, $t_6(Naion)$ are updated to ``China''.  $r$ also verifies $t_2(Nation), t_3(Capital), t_4(Nation)$, as well as the values of $Nation$ and $Capital$ in $t_1,t_5,t_7,$ and $t_8$. $\qed$
\end{example}

\begin{definition}
Applying $R=\{r_1,\dots,r_c\}$ to $t$, denoted as $t \rightarrow_{R} \acute{t}$, is to retrieve a unique final repair $\acute t$, $\forall t \in D$, after a series of modifications as $t\rightarrow_{r_1} t_1 \dots \rightarrow_{r_c} t_c$, and whatever is the order in which the rules in $R$ are appropriately applied.
\end{definition}

\begin{definition}
$t$ has a unique repair by $R$ if there is only one $\acute t$ such that $t \rightarrow_{R} \acute{t}$.
\end{definition}

To guarantee a unique final repair of each $t_i \in D$ by applying $R$, 
\emph{verified attributes} $VA_i$ are defined whose values can not be updated by $R$. Then, we add an additional condition to apply a rule $r_k \in R$ to $t_i$, denoted as, $y_k \not \in VA_i \parallel X_k \not \subset VA_i$, which imposes that $y_k$ is not an attribute in $VA_i$ or $X_k$ is not a subset of $VA_i$.

\section{Weighted Matching Rectifying Rule Discovery}
\label{section:Rule Discovery}
In this section, we design our proposed rule discovery algorithm, WMRRD. First, we define the rule discovery problem in the data repairing context. Next, we develop WMRRD algorithm to create and weight rules automatically from dirty data in-hand. Finally, we study the time complexity of this algorithm.

\begin{problem}
Given a data set $D$ over a schema $S$ and a set $\Sigma$ of functional dependencies over $D$, the rule discovery problem is to discover a WMRR set $R$ automatically from the data $D$ based on $\Sigma$ without need of any external data source.
\end{problem}

Since every weighted matching rectifying rule is built on semantically correlated attributes, our rule discovery algorithm exposes the violations of given data functional dependencies and creates rules based on the assumption that the correct value of an attribute has a higher frequency than its wrong values (Assumption 1).

In our algorithm WMRRD (shown in Algorithm 1) and their procedures (shown in Algorithm 1 cont.), for each FD $\varphi_j:X_j \rightarrow y_j$, the discovering process follows the next steps to create $R$.

\begin{table}[t]
\label{Algorithm: WMRRD}
\begin{tabular*}{\textwidth}{rl}
\hline\noalign{\smallskip}
\multicolumn{2}{l}{Algorithm 1 WMRRD}\\
\noalign{\smallskip}\hline\noalign{\smallskip}
\multicolumn{2}{l}{Input: a dirty dataset $D$, a set of FDs $\Sigma, \theta$} \\
\multicolumn{2}{l}{Output: a WMRR set $R$} \\
1:&begin\\
2:& $R \leftarrow \phi $\\
3:&for each FD $\varphi_j:X_j \rightarrow y_j \in \Sigma$ do\\
4:&\quad  $VP_j \leftarrow$ getVerticalProjection($\varphi_j,D$)\\
5:&\quad  $XY_j \leftarrow$ getHorizontalProjection($VP_j$)\\
6:&\quad for each $(P_i(X_j),P_i(y_j)) \in XY_j$ do\\
7:&\quad \quad if $|P_i(y_j)| > 1$ then\\
8:&\quad \quad \quad $DP_i(X_j) \leftarrow P_i(X_j)$\\
9:&\quad \quad \quad $cp_i(y_j) \leftarrow argmax \{freq(p_k(y_j))|p_k(y_j) \in P_i(y_j)\}$\\
10:&\quad \quad \quad $WP_i(y_j) \leftarrow \{p_k(y_j) | freq(p_k(y_j)) <freq(cp_i(y_j)), \forall p_k(y_j) \in P_i(y_j)\} $\\
11:&\quad \quad \quad $r_{ij} \leftarrow [X_j \approx_X DP_i(X_j)] \land [y_j \in WP_i(Y_j)] \Rightarrow [DP_i(X_j)] \land [cp_i(y_j)$]\\
12:&\quad \quad \quad \(\displaystyle  w_1(r_{ij})=\frac{|DP_i(X_j) \cup cp_i(y_j)|}{|DP_i(X_j)|}\); \(\displaystyle  w_2(r_{ij})=\frac{|DP_i(X_j) \cup cp_i(y_j)|}{|D|}\)\\
13:&\quad \quad \quad if $w_1(r_{ij}) > \theta $\\
14:&\quad \quad \quad \quad  $ R \leftarrow R \cup \{r_{ij}\}$\\
15:&\quad \quad \quad end if\\
16:&\quad \quad end if\\
17:&\quad end for\\
18:&end for\\
19:&end\\
\noalign{\smallskip}\hline
\end{tabular*}
\end{table}

\emph{Step 1} (lines 3--5). We build a hash map $XY_j$ to index data tuples of $X_j \cup \{y_j\}$, which are held in $VP_j$ by \textbf{getVerticalProjection} procedure. For this end, we partition tuples of $VP_j$ according to $X_j$ patterns using \textbf{getHorizontalProjection} procedure, where each part $d_i$ composes an element in $XY_j$; $d_i$ is built on a specific pattern $P_i(X_j)$ and a set of different $y_j$ patterns, $P_i(y_j)$. Each pattern in $P_i(y_j)$ is attached with its frequency $freq_{ij}$ in $d_i$. Then, a hash map $XY_j$ has the following structure: $\{(P_1(X_j), P_1(y_j)), \dots (P_{n_j}(X_j),$ $P_{n_j}(y_j))\}$, such that $P_i(y_j)=\{(p_1(y_j), freq_{1j}), \dots (p_{m_i}(y_j), freq_{m_ij})\}$, where $n_j$ is the number of distinct $X_j$ patterns in $VP_j$, and $m_ij$ is the number of distinct $y_j$ patterns in $d_i$.


\emph{Step 2} (lines 6--10). We classify $y_j$ patterns in each part $d_i$ according to their frequency and based on Assumption 1. Thus, the value with maximum frequency is correct and other values are wrong.

\begin{table}[t]
\label{Algorithm: WMRRD procedures}
\begin{tabular*}{\textwidth}{rl}
\hline\noalign{\smallskip}
\multicolumn{2}{l}{Algorithm 1 WMRRD cont.}\\
\noalign{\smallskip}\hline\noalign{\smallskip}
20:&Procedure \textbf{getVerticalProjection}($\varphi_j$,$D$)\\
21:&\quad $VP_j\leftarrow \phi$\\
22:&\quad for each $a_i \in A$ do\\
23:&\quad \quad if $a_i \in X_j$ or $a_i=y_j$ then \\
24:&\quad \quad \quad $VP_j\leftarrow VP_j \cup \{a_i\}$\\
25:&\quad \quad end if\\
26:&\quad return $VP_j$\\
27:&end procedure\\
\\
28:&Procedure \textbf{getHorizontalProjection}($VP_j$)\\
29:&\quad $XY_j\leftarrow \phi$\\
30:&\quad for each $t_i$ in $VP_j$ do\\
31:&\quad \quad $P(X_j) \leftarrow t_i(X_j)$\\
32:&\quad \quad $P(y_j)\leftarrow \phi$\\
33:&\quad \quad if $P(X_j)$ is a key in $XY_j$ then\\
34:&\quad \quad \quad $P(y_j)\leftarrow \{P(y_j) |(P(X_j),P(y_j)) \in XY_j\}$ \\
35:&\quad \quad end if \\
36:&\quad \quad if $t_i(y_j)\in P(y_j)$ then\\
37:&\quad \quad \quad add 1 to $freq_{ij}$\\
38:&\quad \quad else \\
39:&\quad\quad \quad $freq_{ij}\leftarrow$ 1 \\
40:&\quad \quad \quad $P(y_j)\leftarrow P(y_j) \cup \{(t_i(y_j),freq_{ij})\}$ \\
41:&\quad\quad end if\\
42:&\quad \quad put $(P(X_j),P(y_j))$ into $XY_j$\\
43:&\quad end for\\
44:&\quad return $XY_j$\\
45:&end procedure\\
\noalign{\smallskip}\hline
\end{tabular*}
\end{table}

\emph{Step 3} (lines 11--19). A rule $r_{ij}$ is created as follows: (1) the director pattern is $P_i(X_j)$, the correct pattern is $y_j$' pattern with the maximum frequency, and (3) the wrong patterns are $y_j$' patterns with frequencies less than maximum. Two weights are calculated for $r_{ij}$ based on Eq. (\ref{weight1 definition}) and Eq. (\ref{weight2 definition}). $r_{ij}$ is adopted if $w_1(r_{ij})$ is no less than a given threshold $\theta$.

\begin{example}
Consider the data set in Table \ref{table:motivated example table}. A hash map $XY$ is built w.r.t. $fd$ in Example \ref{example:motivated example}, to index data tuples $\{Nation$ $\cup$ $Capital\}$ as: $XY=$ \{(``China'', \{(``Beijing'' ,3), (``Hongkong'', 1), (``Shanghai'', 1)\}), (``Chine'', \{(``Beijing'', 1)\}), (``Chiena'', \{(``Hongkong'', 1)\})\}. Suppose $\theta=0.6$, a matching rectifying rule $r$ in Example \ref{example:Matching rectifying rules} is created and adopted. 
\end{example}

In contrast to all other existing data repairing rules \cite{fan2012towards,interlandi2015proof,wang2014towards}, weighted matching rectifying rules are full-automatically discovered by Algorithm 1, from the dirty data in-hand and without external master data. 

\textbf{Complexity} The outer loop (line 3) iterates $|\Sigma|$ times. Vertical projection (line 4) runs in time linear to $|A(D)|$, .i.e, the number of data attributes. Horizontal Projection (line 5) runs in time $|D|*|XY_j|$, where $|XY_j|$ is the number of distinct frequent $X_j$ patterns in $D$. Then, line 5 in the worst case runs in $|D|^2$ times. The inner loop (lines 6--17) runs in time $\sum_{i=1}^{|XY_j|} |P_i(y_j)|$ which equals $|D|$ in the worst case. Accordingly, the total time complexity of Algorithm 1 is $O(|\Sigma|.(|A|+|D|^2+|D|))$.

Although the time complexity of our rule discovery algorithm in the worst case is quadratic in number of tuples, data sets often have many frequent $X_j$ and $y_j$ patterns in practice, so using the hash map $XY_j$ can decrease the time complexity to be approximately linear as we see later in the experiments.

\section{Fundamental Problems}
\label{section:Fundamental Problems}
\subsection{Termination}
One regular problem for rule-based data repairing methods is termination. Given a data set $D$ and a set of rules $R$, the \emph{termination problem} is to define whether each repairing process on $D$ will end based on $R$.

Indeed, it is easy to ensure that the repairing process ends by applying a WMRR set to each tuple. Let $t \in D$ be a data tuple, and $R$ be a WMRR set. According to the rule semantics in Sect. \ref{subSection:Rule Semantics}, repairing each tuple $t$ based on $R$ is a series of modifications which ends up with a final repair $\acute t$.

\subsection{Consistency}
Given a WMRR set $R$ over a data set $D$, the \emph{consistency problem} is to define whether $R$ is a consistent set, i.e., whether applying $R$ outputs a unique repair for all different applicable rules order.

$R$ is consistent set iff $r_i$ and $r_j$ are consistent $\forall r_i, r_j \in R$ \cite{wang2014towards}.

\begin{theorem}
\label{Theorem: consistency problem}
The consistency problem of WMRRs is PTIME.
\end{theorem}

We prove Theorem \ref{Theorem: consistency problem} by developing a PTIME algorithm in Sect. \ref{section:Rule Inconsistency Resolution}, which checks rule consistency and also solves rule inconsistency.

\subsection{Determinism}
The \emph{Determinism problem} is to define whether all possible terminating repairing processes lead to a unique repair.

According to the consistency condition and the rule semantics in Sect. \ref{subSection:Rule Semantics}, a unique final repair $\acute t$ is retrieved by applying a consistent set of WMRRs to each tuple $t \in D$. Thus, repairing $D$ is deterministic.

\subsection{Implication}
Given a consistent set $R$ of WMRRs, and another rule $r \not \in R$, the \emph{implication problem} is to define whether $R$ implies $r$, denoted as $R \models r$.

\begin{definition}
$R \models r$ if (1) $R\cup\{r\}$ is a consistent set, and (2) $\forall t\in D, t \rightarrow_R \acute t \land t \rightarrow_{R\cup\{r\}} \acute t$. (1) means that there is no conflict between $R$ and $r$. (2) means that any data tuple will be rectified uniquely by applying either $R$ or $R \cup\{r\}$, which marks $r$ as an unnecessary rule.
\end{definition}

\begin{theorem}
In general, the implication problem of WMRRs is coNP-complete, but it is PTIME when the data set is fixed \cite{wang2014towards}.
\end{theorem}

\section{Rule Inconsistency Resolution}
\label{section:Rule Inconsistency Resolution}
\begin{definition}
Given a WMRR set $R$ over $D$, and two different rules $r_i, r_j \in R$. Based on the consistency problem definition, $r_i$ and $r_j$ are consistent iff $\forall t\in D, t$ is rectified to $\acute t$ either we apply $r_i$ then $r_j$, or $r_j$ then $r_i$.
\end{definition}

We develop an automatic algorithm Inconsis-Res (shown in Algorithm 2) for WMRRs inconsistency resolution, which checks the consistency for each pair of the rules and solve the inconsistency automatically.

\begin{table}[h]
\label{table:Algorithm 2 Inconsis-Res}
\begin{tabular*}{\textwidth}{rl}
\hline\noalign{\smallskip}
\multicolumn{2}{l}{Algorithm 2 Inconsis-Res}\\
\noalign{\smallskip}\hline\noalign{\smallskip}
\multicolumn{2}{l}{Input: a WMRR set $R$} \\
\multicolumn{2}{l}{Output: a consistent set  $\acute{R}$}\\
1:&begin\\
2:&$\acute{R} \leftarrow$ $R$\\
3:&for each $r_i, r_j \in  R$ do\\
4:&\quad $consis\leftarrow True$\\
5:&\quad if $X_i \cap X_j= \phi$ or $DP_i(X_i \cap X_j)\approx_{X} DP_j(X_i \cap X_j)$ then\\
6:&\quad \quad if $y_i=y_j$ then\\
7:&\quad \quad \quad if $cp_i(y_i)\neq cp_j(y_i)$ and $WP_i(y_i) \cap WP_j(y_i) \neq \phi$ then\\
8:&\quad \quad \quad \quad $consis\leftarrow False$\\
9:&\quad \quad \quad  end if\\
10:&\quad \quad else if $y_j \in X_i$ and $y_i \notin X_j$ and $DP_i(y_j) \in WP_j(y_j)$ then\\
11:&\quad \quad \quad  $consis\leftarrow False$\\
12:&\quad \quad else if $y_i \in X_j$ and $y_j \notin X_i$ and $DP_j(y_i) \in WP_i(y_i)$ then\\
13:&\quad \quad \quad  $consis\leftarrow False$\\
14:&\quad \quad else if $y_i \in X_j$ and $y_j \in X_i$ and $DP_j(y_i) \in WP_i(y_i)$  and $DP_i(y_j) \in WP_j(y_j)$ then\\
15:&\quad \quad \quad  $consis\leftarrow False$\\
16:&\quad \quad end if\\
17:&\quad end if\\
18:&\quad if $\neg consis$ then\\
19:&\quad \quad $r \leftarrow minarg\{w_1(r_i),w_1(r_j)\}$\\
20:&\quad \quad $\acute{R}\leftarrow \acute{R}\setminus \{r\}$\\
21:&\quad end if \\
22:&end for\\
23:&end\\
\noalign{\smallskip}\hline
\end{tabular*}
\end{table}
$\forall r_i,r_j \in R$, as follows:

$r_i: [X_i \approx_X DP_i(X_i)]\land [y_i \in WP_i(y_i)]\Rightarrow [DP_i(X_i)] \land [cp_i(y_i)]$.

$r_j: [X_j \approx_X DP_j(X_j)]\land [y_j \in WP_j(y_j)]\Rightarrow [DP_j(X_j)] \land [cp_j(y_j)]$.

First, $r_i$ and $r_j$ are checked.  If both rules have different $X$ attributes or similar direct patterns for the same $X$ attributes, they both can be matched by $t$ (lines 3--5). Therefore, $r_i$ and $r_j$ are considered inconsistent in four conditions:

\begin{enumerate}[label={(\arabic*)}]
\item When $y_i=y_j$. If the rules share wrong patterns without the same correct pattern (lines 6--9).

\item When $y_i \neq y_j$, $y_j \in X_i$, and $y_i \not \in X_j$. If the correct pattern of $y_j$ in $r_i$ is wrong in $r_j$ (lines 10,11). Note for a matching tuple $t$, if $r_i$ is applied first, $t(y_j)$ is correct. But, if $r_j$ is applied first, $t(y_j)$ will be modified.

\item When $y_i \neq y_j$, $y_i \in X_j$, and $y_j \not \in X_i$. If the correct pattern of $y_i$ in $r_j$ is wrong in $r_i$ (lines 12,13).

\item When $y_i \neq y_j$, $y_i \in X_j$, and $y_j \in X_i$. If the correct pattern of $y_j$ in $r_i$ is wrong in $r_j$, and the correct pattern of $y_i$ in $r_j$ is wrong in $r_i$ (lines 14,15).
\end{enumerate}
Second, if $r_i$ and $r_j$ are inconsistent, the rule with less confidence $w_1$ is excluded (lines 18--21).

In contrast to the existing data repairing rules, such as fixing rules \cite{wang2014towards}, where experts are required to resolve the inconsistency, we resolve this problem for WMRRs automatically with keeping high-quality rules.

\emph{Complexity} Since Algorithm 2 checks each pair of rules, its time complexity is $O({|R|}^2)$, where $|R|$ is the rule set size, i.e., the number of rules. However, the algorithm scales better in our experiments.

\section{WMRR-based Data Repairing}
\label{section:WMRR-based Data Repairing}
In this section, we present our data repairing algorithm based on weighted matching rectifying rules, WMRR-DR. First, we define WMRR-based data repairing problem. Then, we develop WMRR-DR algorithm and explain the repairing process of this algorithm. Finally, we study the time complexity of the algorithm.

\begin{problem}
\label{Problem: WMRR-based Data Repairing}
Given a data set $D$ over a schema $S$ and a consistent set $R$ of WMRRs over $D$, WMRR-based data repairing problem is to retrieve a valid and unique repair $\acute{D}$ of $D$ by detecting errors in $D$ and rectify the detected errors uniquely, dependably and automatically without user verifications.
\end{problem}

To efficiently use $R$ in the repairing process, we index it as a hash map $IR$ in order to efficiently determine the candidate rules $CR$ for each tuple, as we see in the next steps. $IR$ is a mapping from an attribute-value pair $p(a,v)$ to a WMRR set $R_p$, such that $\forall r_k \in R_p; r_k$ matches $p$, i.e., $a \in X_k \land DP_k(a)=v$.

\begin{table}[h]
\label{table:Algorithm3 WMRR-DR}
\begin{tabular*}{\textwidth}{rl}
\hline\noalign{\smallskip}
\multicolumn{2}{l}{Algorithm 3 WMRR-DR}\\
\noalign{\smallskip}\hline\noalign{\smallskip}
\multicolumn{2}{l}{Input: a dirty data set $D$, a set of FDs $\Sigma, IR$} \\
\multicolumn{2}{l}{Output: a rectified data set $\acute{D}$ }\\
1:&begin\\
2:&$\acute{D} \leftarrow \phi$ \\
3:&for each $t_i$ in $D$ do\\
4:&\quad $\acute t_i\leftarrow t_i$\\
5:&\quad $CR_i \leftarrow \phi$, $VA_i \leftarrow \phi$\\
6:&\quad for each attribute-value pair $p \in t_i$ do \\
7:&\quad \quad $CR_i \leftarrow CR_i \cup IR_p$\\
8:&\quad \quad if $IR_p = \phi $ then\\
9:&\quad \quad \quad $CR_i \leftarrow CR_i \cup IR_{\approx p}$\\
10:&\quad \quad end if\\
11:&\quad end for\\
12:&\quad for each $\varphi_j \in \Sigma$ do\\
13:&\quad \quad $R_{\varphi_{j}}\leftarrow$ getFdRules($\varphi_{j}$,$CR_i$)\\
14:&\quad \quad if $R_{\varphi_{j}} \neq \phi$ then\\
15:&\quad \quad \quad $R(t_i)\leftarrow$ findMatchingRules($R_{\varphi_{j}}$,$t_i$)\\
16:&\quad \quad \quad $\acute R(t_i)\leftarrow$ filterMatchingRules($R(t_i)$,$t_i$)\\
17:&\quad \quad \quad for each $r_k \in \acute R(t_i)$ do\\
18:&\quad \quad \quad \quad if $y_k \notin VA_i$ then \\
19:&\quad \quad \quad \quad \quad $\acute t_i(y_j)\leftarrow cp_k(y_j)$\\
20:&\quad \quad \quad \quad \quad $VA \leftarrow VA \cup \{y_j\}$\\
21:&\quad \quad \quad \quad end if \\
22:&\quad \quad \quad \quad if $X_j \not \subset VA_i$ then \\
23:&\quad \quad \quad \quad \quad $\acute t_i(X_j)\leftarrow DP_k(X_j)$\\
24:&\quad \quad \quad \quad \quad $VA \leftarrow VA  \cup X_j$ \\
25:&\quad \quad \quad \quad end if \\
26:&\quad \quad \quad end for\\
27:&\quad \quad end if\\
28:&\quad end for\\
29:&\quad  $\acute D\leftarrow \acute D \cup \{\acute t_i\}$\\
30:&end for\\
31:&end\\
\noalign{\smallskip}\hline
\end{tabular*}
\end{table}

Our algorithm WMRR-DR (Shown in Algorithm 3) addresses Problem \ref{Problem: WMRR-based Data Repairing} by discovering a unique and valid repair $\acute t_i$ for each tuple $t_i \in D$, using two procedures (shown in Algorithm 3 cont.) as follows.

\emph{Step I} (lines 3-11). A candidate rule set $CR_i$ is identified by detecting rules of $IR$ that exactly match a pair $p$ in $t_i$, called $IR_p$. When no rules are founded, $IR_{\approx p}$ is detected as the rules of $IR$ that similarly match $p$, i.e., $a \in X_k \land DP_k(a)\approx_a v$ based on Eq. (\ref{Eq.Similarity metric}) and Eq. (\ref{Eq.Similarity function}), Sect. \ref{subSection:Rule Syntax}.

\emph{Step II} (lines 12-16). To find matching rules: (1) $CR_i$ is classified based on FDs where $\forall r_k \in R_{\varphi_{j}}; X_k=X_j \land y_k=y_j$. (2) A matching rule set $R(t_i)$ is identified by \textbf{findMatchingRules} procedure. (3) $R(t_i)$ is filtered to $\acute R(t_i)$ by \textbf{filterMatchingRules} procedure, in order to assure the correctness of the director pattern of the applied rules. $\acute R(t_i)$ holds the rules with the minimum distance to $t_i$, where this distance is computed based on Eq. (\ref{eq.rule-tuple distance:1}) and Eq. (\ref{eq.rule-tuple distance:2}). If $\acute R(t_i)$ has more than one rule, some dirty rules possibly exist, so $\acute R(t_i)$ is filtered again keeping the rules with the maximum $w_2$ based on Assumption 1.

\begin{definition} The distance between a rule $r_k$ and a tuple $t_i$ is defined as follows when $t_i$ matches $r_k$:
\begin{align}
  dis(r_k,t_i) &= dis(DP_k(X_k),t_i(X_k)),& \label{eq.rule-tuple distance:1}\\
  dis(DP_k(X_k),t_i(X_k)) &= \sum_{n=1}^{|X|} sim(DP_k(x_n), t_i(x_n)),&\label{eq.rule-tuple distance:2}
\end{align}
where $sim(DP_k(x_n), t_i(x_n))$ is defined in Eq. (\ref{Eq.Similarity function}) Sect. \ref{subSection:Rule Syntax}.
\end{definition}

\emph{Step III} (lines 17-31). For each $r_k \in \acute R(t_i)$ that can be applied to $t_i$, $t_i$ is updates and the verified attributes $VA_i$ are extended accordingly.

\begin{table}
\label{table:Procedure of Algorithm 3}
\begin{tabular*}{\textwidth}{rl}
\hline\noalign{\smallskip}
\multicolumn{2}{l}{Algorithm 3 WMRR-ER cont.}\\
\noalign{\smallskip}\hline\noalign{\smallskip}
32:&Procedure \textbf{findMatchingRules}($R_{\varphi_{j}}$,$t_i$)\\
33:&\quad $R(t_{i})\leftarrow \phi$\\
34:&\quad for each $r_k$ in $R_{\varphi_{j}}$ do\\
35:&\quad \quad if $t_i(y_j)=cp_k(y_j)$ or $t_i(y_j) \in WP_k(y_j)$ then \\
36:&\quad \quad \quad $R(t_{i})\leftarrow R(t_{i}) \cup \{r_k\}$\\
37:&\quad \quad end if\\
38:&\quad return $R(t_{i})$\\
39:&end procedure\\
\\
40:&Procedure \textbf{filterMatchingRules}($R(t_i)$,$t_i$)\\
41:&\quad $\acute R(t_i)\leftarrow argmin\{dis(r_k,t_i)| r_k \in R(t_i)\}$\\
42:&\quad if $|\acute R(t_i)|>1$ then\\
43:&\quad \quad $\acute R(t_i)\leftarrow argmax\{w_2(r_k) | r_k \in \acute R(t_i)\}$\\
44:&\quad end if\\
45:&\quad return $\acute R(t_i)$\\
46:&end procedure\\
\noalign{\smallskip}\hline
\end{tabular*}
\end{table}

The following example explains the importance of filtering rules based on the distance criterion followed by the weight criterion.
\begin{example}
Consider the data set in Table \ref{table:motivated example table} and the rule in Example \ref{example:Matching rectifying rules} as $r_1$. Suppose another rule with a wrong directory pattern as: $r_2: ((Nation \approx$ ``Chena''$),(Capital \in $ \{``Hongkong''\}) $\Rightarrow ($ ``China''$ ) \land ($ ``Beijing''$))$. To repair $t_2$ as example, $R(t_2)=\{r_1,r_2\}$. $\acute R(t_2)=\{r_1\}$ since $dis(r_1,t_2)< dis(r_2,t_2)$. Then, $t_2(Capital)$ is rectified to ``Beijing''. To repair $t_6$ as another example, $R(t_6)=\{r_1,r_2\}$. First, $\acute R(t_6)=\{r_1,r_2\}$. Based on Assumption 1, $w_2(r_1)>w_2(r_2)$ where $DP_2(Nation)$ is wrong. Then, $\acute R(t_6)$ is updated to $\{r_1\}$. Accordingly, $t_6(Nation)$ is rectified  to ``China'', and $t_6(Capital)$ is rectified  to ``Beijing''.
\end{example}

\textbf{Complexity} The outer loop (lines 4--31) iterates $|D|$ times to repair all data where each iteration rectifies one tuple. The first inner loop (lines 6--11) runs in time linear to $|IR|$ which in the worst case equals to $|R|$. The second inner loop (lines 12--28) runs in time linear to $|\Sigma|$ since the size of $|CR_i|, |R_{\varphi_{j}}|, |R(t_i)|$, and $|\acute R(t_i)|$ are indeed small enough to consider as constants. The number of FDs $|\Sigma|$ is also small compared with the number of rules $|R|$. Accordingly, the total time complexity of Algorithm 3 is $O(|D|.|R|)$.

\section{Experimental Results}
\label{section:Experiment Results}
In this section, we discuss our extensive experiments to evaluate our rule-based data repairing method including WMRRD, Inconsis-Res and WMRR-DR algorithms where WMRR-DR repairs data errors based on a consistent set of WMRRs that were discovered by WMRRG and checked by Inconsis-Res. First, we evaluate the effectiveness of our data repairing method. Then, we study the effect of threshold $\theta$ on the accuracy of data repairing and the number of discovered rules. After that, we check the effect of typo rate on the number of discovered rules and how varying the number of rules affects the data repairing accuracy. Finally, we study the efficiency of our three algorithms.

\subsection{Experiments Context}
We conducted the experiments on 3.2GHZ Intel(R) core(TM)i5 processor with 4GB RAM, using Microsoft Windows 10, and all algorithms were implemented by Java.

\textbf{Data Sets.} We performed our experiments on both real-life and synthetic data. (1) Hospital\footnote{http://www.hospitalcompare.hhs.gov/} data set (HOSP) is a public data set provided by USA department of Health and Human Service. It consists of 115K tuples with 17 attributes, and 24 FDs. (2) Address\footnote{http://www.cs.utexas.edu/users/ml/riddle/data.html} data set (UIS) is a synthetic data set generated by the UIS data set generator. It consists of 15K tuples with 11 attributes, and 18 FDs. Table \ref{table:FDs of Datasets} shows the functional dependencies over each data set.

\textbf{Noise.} We added two kinds of errors to the attributes on which FDs were defined: (1) typos; (2) active domain errors where a value in a tuple is changed to a different value from other tuples. The clean data sets were used as ground truth. Errors were generated by adding noise with a specific rate (10\% by default).

\begin{table}
  \caption{FDs for Data Sets}
    \begin{tabular}{l}
    \hline\noalign{\smallskip}
    \textbf{FDs for Address} \\
     \noalign{\smallskip}\hline\noalign{\smallskip}
    SSN $\rightarrow$ Fname, Minit, Lname, Stnum, Stadd, Apt, City, State,ZIP \\
    Fname, Minit, Lname $\rightarrow$ SSN, Stnum, Stadd, Apt, City, State, ZIP \\
    ZIP $\rightarrow$ State, City \\
     \noalign{\smallskip}\hline\noalign{\smallskip}
    \textbf{FDs for Hospital} \\
     \noalign{\smallskip}\hline\noalign{\smallskip}
    PN $\rightarrow$ HN, Addr$_1$, Addr$_2$,Addr$_3$, City, State, ZIP,County, Phn, HT, HO, ES \\
    Phn $\rightarrow$ ZIP, City, State, Addr$_1$, Addr$_2$,Addr$_3$ \\
    MC $\rightarrow$ MN, Condition \\
    PN,MC $\rightarrow$ StateAvg \\
    State,MC $\rightarrow$ StateAvg \\
    ZIP $\rightarrow$ State, City \\
     \noalign{\smallskip}\hline
    \end{tabular}%
    \label{table:FDs of Datasets}
\end{table}%

\textbf{Algorithms}. We implemented the three proposed algorithms: (1) WMRRD: the rule discovery algorithm (Sect. \ref{section:Rule Discovery}); (2) Inconsis-Res: the inconsistency resolution algorithm for the discovered rules (Sect. \ref{section:Rule Inconsistency Resolution}); (3) WMRR-DR: the data repairing algorithm based on the discovered consistent rules (Sect. \ref{section:WMRR-based Data Repairing}). For comparison, we implemented the dependable and automatic data repairing method, FR-DR, based on fixing rules that were provided by experts \cite{wang2014towards}.

\textbf{Measuring Quality.} For a fair comparison with the state-of-the-art FR-DR method, we used the accuracy measures, $recall$, $precision$, and $f-measure$: $precision$ is the ratio of the correctly rectified attribute values to all rectified attribute values, $recall$ is the ratio of the correctly rectified attribute values to all wrong attribute values. $precision$ assess correctness of repairing while $recall$ assess completeness of repairing, and $f-measure$ is the harmonic mean of precision and recall, which is defined as follows.
\begin{equation}
f-measure=\frac{2\times precision\times recall}{percision + recall}
\end{equation}

\subsection{Effectiveness Comparison}
\label{subsection:Effectiveness Comparison}
In the first experiment, we compared the effectiveness of our repairing method, WMRR-DR with FR-DR on both data sets. The comparison results are shown in Table \ref{table:Accuracy for Address data} for UIS and Table \ref{table: Accuracy for Hospital data} for HOSP, where we fixed the noise rate at 10\%, varied the typo rate from 0\% to 100\%, and reported the recall, the precision and the number of repairs (\#Repair). We set the threshold $\theta$ = 0.6 by default and studied its effect next in Sect. \ref{subsection: Effect of Threshold}. Both tables show that our method outperforms FR-DR in recall for all adopted typo rates, with maintaining 100\% of precision. This is due to the fact that our method rectifies correctly a greater number of errors than FR-DR, since WMRRs depend on similarity matching to detect and repair more errors. Furthermore, WMRRs are built on the data that is most likely to be correct, and weighted to ensure their quality.
\begin{table}
  \caption{Repairing Accuracy Comparison on Address data}
    \begin{tabular}{lllllll}
    \hline\noalign{\smallskip}
    {\textbf{Typo-rate}} & \multicolumn{3}{l}{\textbf{FR-DR}} & \multicolumn{3}{l}{\textbf{WMRR-DR}} \\
& \textbf{\#Repair} & \textbf{Recall} & \textbf{Precision} & \textbf{\#Repair} & \textbf{Recall} & \textbf{Precision} \\
    \noalign{\smallskip}\hline\noalign{\smallskip}
    0     & 1     & 0.0001 & 1     & 692   & \textbf{0.0689} & \textbf{1} \\
    0.1   & 4     & 0.0004 & 1     & 676   & \textbf{0.0617} & \textbf{1} \\
    0.2   & 18    & 0.0015 & 1     & 708   & \textbf{0.0597} & \textbf{1} \\
    0.3   & 21    & 0.0017 & 1     & 684   & \textbf{0.0546} & \textbf{1} \\
    0.4   & 29    & 0.0022 & 1     & 755   & \textbf{0.0562} & \textbf{1} \\
    0.5   & 33    & 0.0023 & 1     & 672   & \textbf{0.0469} & \textbf{1} \\
    0.6   & 35    & 0.0023 & 1     & 759   & \textbf{0.0501} & \textbf{1} \\
    0.7   & 51    & 0.0032 & 1     & 701   & \textbf{0.0443} & \textbf{1} \\
    0.8   & 39    & 0.0023 & 1     & 732   & \textbf{0.0437} & \textbf{1} \\
    0.9   & 49    & 0.0028 & 1     & 709   & \textbf{0.0406} & \textbf{1} \\
    1     & 58    & 0.0032 & 1     & 773   & \textbf{0.0422} & \textbf{1} \\
    \noalign{\smallskip}\hline
    \end{tabular}
    \label{table:Accuracy for Address data}
\end{table}

\begin{table}
  \caption{Repairing Accuracy Comparison on Hospital data}
    \begin{tabular}{lllllll}
    \hline\noalign{\smallskip}
    {\textbf{Typo-rate}} & \multicolumn{3}{l}{\textbf{FR-DR}} & \multicolumn{3}{l}{\textbf{WMRR-DR}} \\
& \textbf{\#Repair} & \textbf{Recall} & \textbf{Precision} & \textbf{\#Repair} & \textbf{Recall} & \textbf{Precision} \\
    \noalign{\smallskip}\hline\noalign{\smallskip}
    0     & 1155  & 0.011 & 1     & 75544 & \textbf{0.69} & \textbf{1} \\
    0.1   & 2998  & 0.026 & 1     & 80544 & \textbf{0.705} & \textbf{1} \\
    0.2   & 4915  & 0.041 & 1     & 86187 & \textbf{0.725} & \textbf{1} \\
    0.3   & 6702  & 0.055 & 1     & 90989 & \textbf{0.74} & \textbf{1} \\
    0.4   & 8461  & 0.066 & 1     & 96371 & \textbf{0.756} & \textbf{1} \\
    0.5   & 10115 & 0.077 & 1     & 101136 & \textbf{0.766} & \textbf{1} \\
    0.6   & 12345 & 0.09  & 1     & 107584 & \textbf{0.786} & \textbf{1} \\
    0.7   & 13948 & 0.099 & 1     & 113050 & \textbf{0.8} & \textbf{1} \\
    0.8   & 15628 & 0.107 & 1     & 118092 & \textbf{0.809} & \textbf{1} \\
    0.9   & 17632 & 0.117 & 1     & 123476 & \textbf{0.82} & \textbf{1} \\
    1     & 18970 & 0.122 & 1     & 127179 & \textbf{0.82} & \textbf{1} \\
    \noalign{\smallskip}\hline
    \end{tabular}
    \label{table: Accuracy for Hospital data}
\end{table}

For the sensitivity to typos, we can observe that the recall increases with the growth of typo rate on HOSP, but it fluctuates on UIS because HOSP has more frequent patterns for each FD than UIS, then the generated typos are more likely to place in these patterns and then detected and rectified.

Since our method has higher recall than FR-DR with the same precision for each adopted typo rate, we measure the improvement of accuracy in term of $avg.f-measure$ on both data sets, as shown in Table \ref{table: Avg-f-measure Comparison}. 
The results show that our method improves the accuracy up to 9.4\% for UIS and up to 73\% for HOSP. These findings verify that our method discovers effective rules and repairs errors based on these rules effectively. In the next experiments, the accuracy will be evaluated using $f-measure$.

\begin{table}[H]
  \caption{$Avg.f-measure$ Comparison}
    \begin{tabular}{lll}
   \hline\noalign{\smallskip}
    $Avg.f-measure$ & \multicolumn{1}{l}{FR-DR} & \multicolumn{1}{l}{WMRR-DR} \\
    \noalign{\smallskip}\hline\noalign{\smallskip}
    Address & 0.004 & \textbf{0.098} \\
    Hospital & 0.14  & \textbf{0.87} \\
     \noalign{\smallskip}\hline
    \end{tabular}
     \label{table: Avg-f-measure Comparison}
\end{table}

\subsection{Effect of Threshold $\theta$}
\label{subsection: Effect of Threshold}
\begin{figure}
       \centering
       \captionsetup{justification=centering}
        \subfloat[\#Rules of UIS]{\label{figure:thresholdRulesUIS}\includegraphics[scale=0.50] {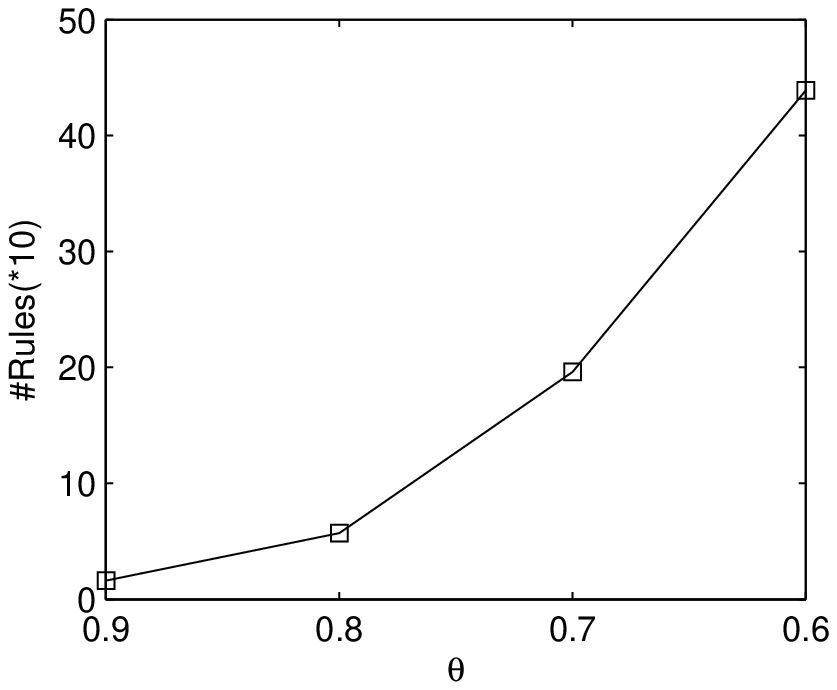}}
        \subfloat[\#Rules of HOSP]{\label{figure:thresholdRulesHOSP}\includegraphics[scale=0.50] {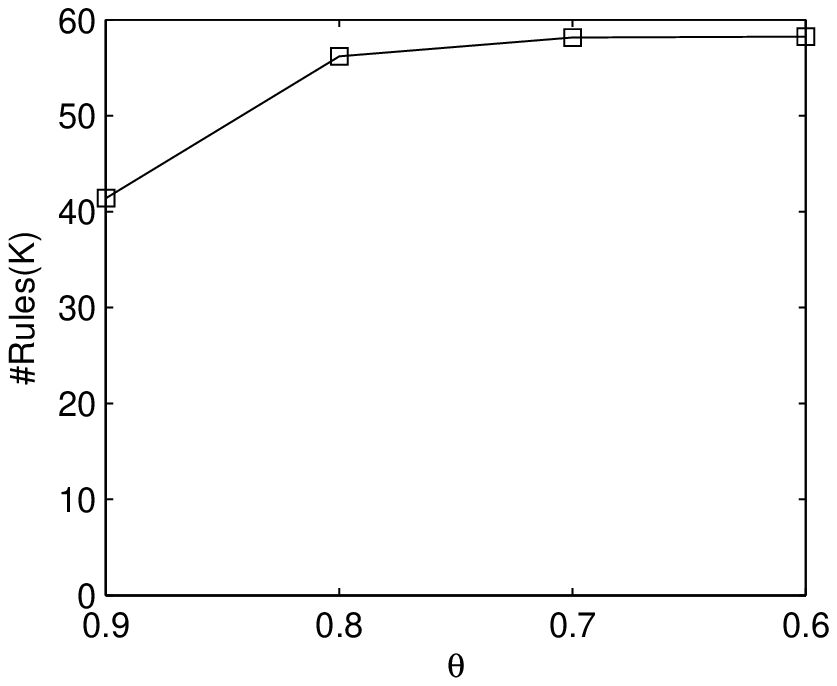}}\\
        \subfloat[UIS Accuracy]{\label{figure:thresholdAccuracyUIS}\includegraphics[scale=0.50] {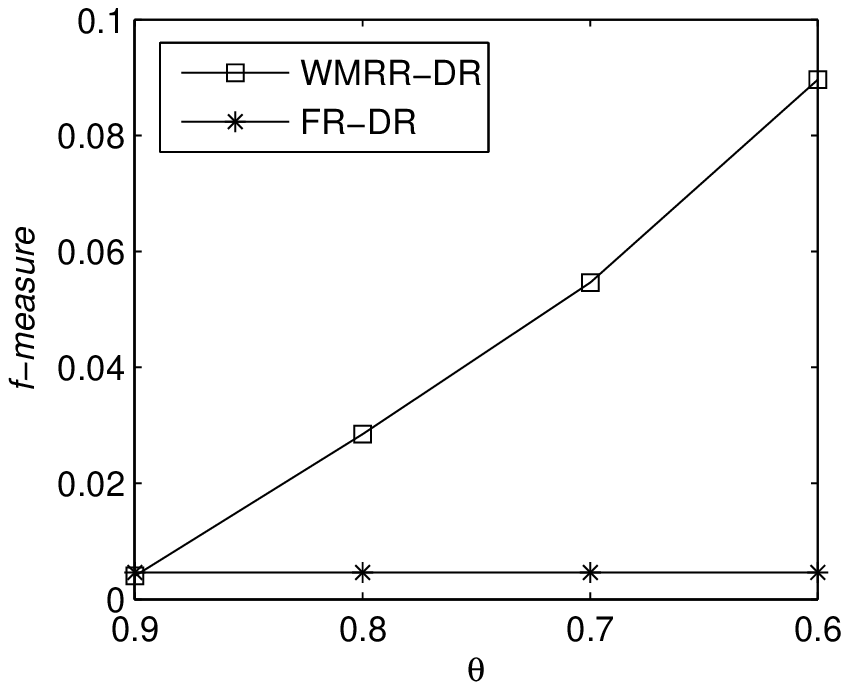}}
        \subfloat[HOSP Accuracy]{\label{figure:thresholdAccuracyHOSP}\includegraphics[scale=0.50] {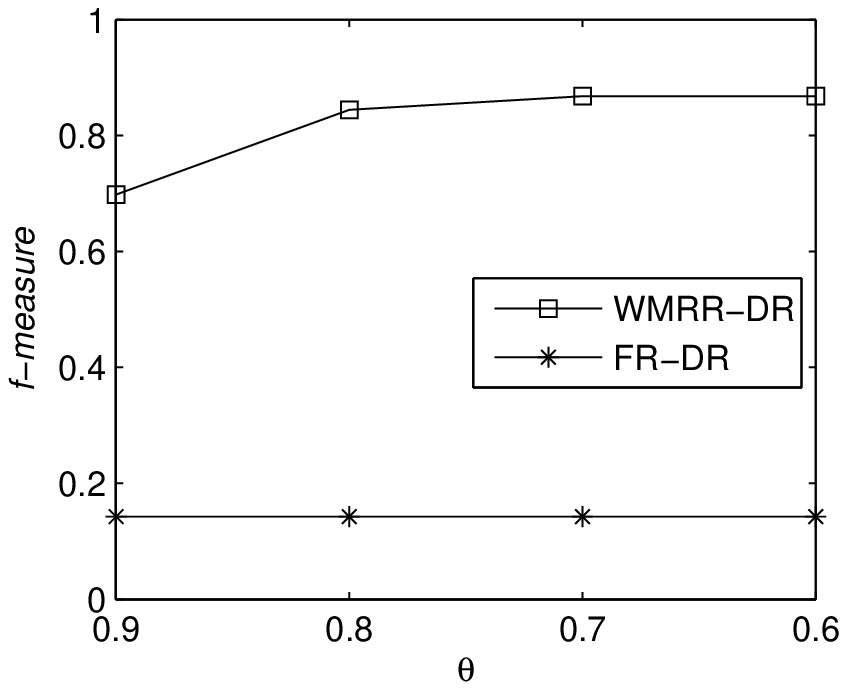}}
        \caption{Effect of threshold $\theta$ on Accuracy and \#Rules}\label{Effect of thresold on Accuracy}
\end{figure}

First, we checked the effect of decreasing the threshold $\theta$ from 0.9 down to 0.6 on the discovered rule set size for the two data sets with typo rate 50\%. Figs. \ref{figure:thresholdRulesUIS} and \ref{figure:thresholdRulesHOSP} report the rule set size, i.e., the number of rules, on UIS and HOSP data sets, respectively. We observe the following: (1) The rule set size increases while decreasing $\theta$ since more rules will be discovered and adopted by WMRRD. (2) The growth of rule set size is greater for UIS than HOSP since the attribute values in UIS are less frequent than they are in HOSP; for example, the rule set size is almost the same for both thresholds 0.7 and 0.6  on HOSP, while the rule set size for $\theta=0.6$ is more than the double for $\theta=0.7$ on UIS.

Then, with the same settings, we studied the accuracy of WMRR-DR for these different thresholds on the two data sets. Figs. \ref{figure:thresholdAccuracyUIS} and \ref{figure:thresholdAccuracyHOSP} report $f-measure$ results on UIS and HOSP, respectively. 
They show that the accuracy of our method increases gradually with the drop of $\theta$, as expected from the growth of rule set size, where the accuracy reaches 87 \% on HOSP for $\theta=0.7$, and it reaches 9\% on UIS when $\theta=0.6$. Moreover, our method outperforms FR-DR significantly in accuracy for all thresholds, except for $\theta=0.9$ on UIS where both methods have the same accuracy since the attribute values in UIS are little frequent. Accordingly, we adopted $\theta=0.6$ for UIS, and $\theta=0.7$ for HOSP in our next experiments.

\subsection{Effect of Typo Rate and Rule Set Size}
\begin{figure}
       \centering
       \captionsetup{justification=centering}
        \subfloat[\#Rules of UIS]{\label{figure:typoRateRulesUIS}\includegraphics[scale=0.5] {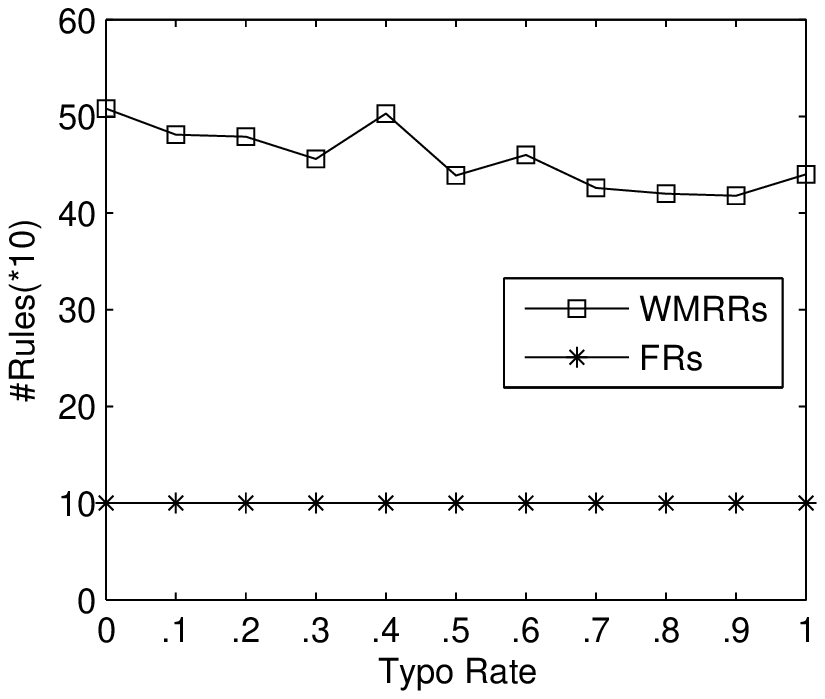}}
        \subfloat[\#Rules of HOSP]{\label{figure:typoRateRulesHOSP}\includegraphics[scale=0.5] {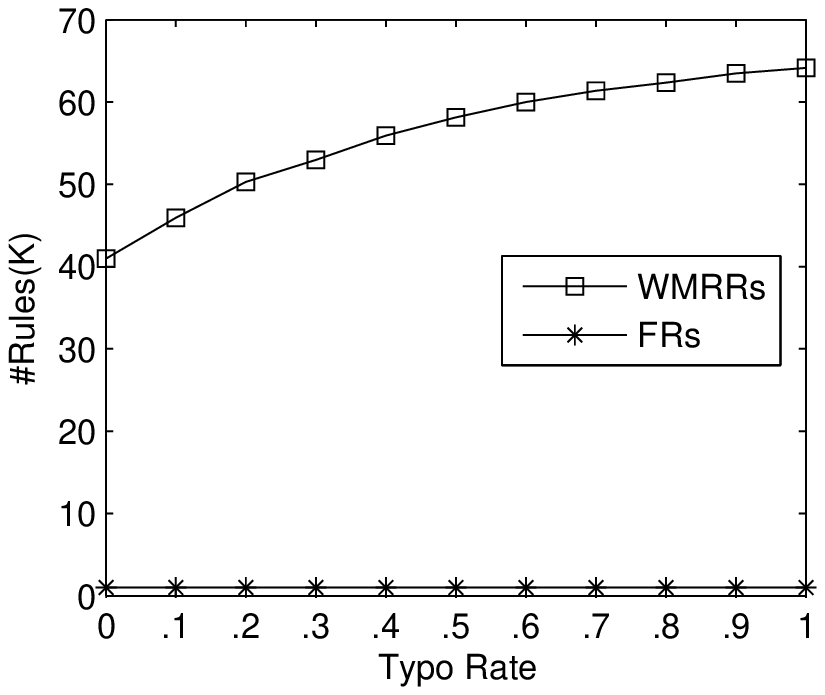}}
        \caption{Effect of Typo-Rate on \#Rules}\label{figure:Effect of Typo-Rate on RulesNum}
\end{figure}

We investigated the number of discovered weighted matching rectifying rules (WMRRs) compared with the number of fixing rules (FRs) with different typo rates. We increased the typo rate from 0\% to 100\% and reported the number of both kinds of rules on UIS and HOSP in Figs. \ref{figure:typoRateRulesUIS} and \ref{figure:typoRateRulesHOSP}, respectively. The results show that more WMRRs are discovered with more typos on HOSP, while the number of WMRRs on UIS changes with a narrow fluctuation, but it often decreases little with the growth of typo rate. 
This change depends on to what extent the patterns of each FD are frequent and how the typos are distributed in these frequent patterns. In opposite, the same number of FRs is used even for different typo rates since they are provided by experts one time. These findings approve the accuracy comparison in Sect. \ref{subsection:Effectiveness Comparison}.

\begin{figure}
       \centering
       \captionsetup{justification=centering}
        \subfloat[UIS Accuracy]{\label{figure:rulesRecallUIS}\includegraphics[scale=0.5] {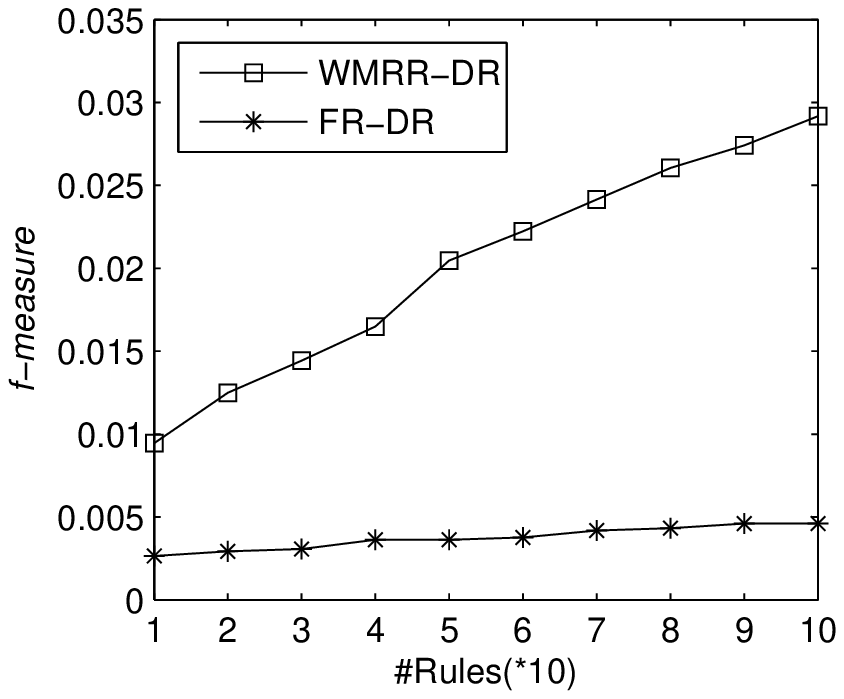}}
        \subfloat[HOSP Accuracy]{\label{figure:rulesRecallHOSP}\includegraphics[scale=0.5] {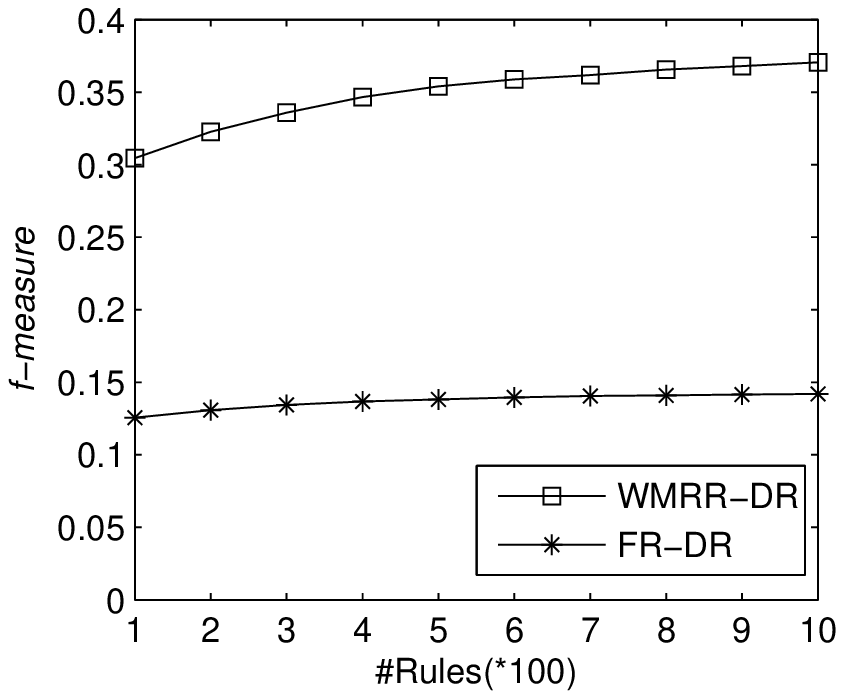}}
        \caption{Effect of \#Rules on Accuracy}\label{figure:Effect of RulesNum on Accuracy}
\end{figure}

For further performance understanding, we also examined the repairing accuracy of our method WMRR-DR compared with FR-DR based on different numbers of rules. 
We increased the number of rules from 10 to 100 for UIS and from 100 to 1000 for HOSP, with typo rate 50\% for both data sets. Figs. \ref{figure:rulesRecallUIS} and \ref{figure:rulesRecallHOSP} report the $f-measure$ on UIS and HOSP, respectively. The results indicate that although both methods can achieve better accuracy by using more rules, our method WMRR-DR is more accurate than FR-DR even by using a little number of discovered rules.


\subsection{Efficiency and Scalability}
On UIS and HOSP, we evaluated the efficiency of WMRRD, and WMRR-DR algorithms by varying the data size, i.e., the number of tuples, and the efficiency of Inconsis-Res by varying the rule set size, i.e., the number of checked rules.

\begin{figure}[h]
       \centering
       \captionsetup{justification=centering}
        \subfloat[Scalability of WMRRD on UIS]{\label{figure:ScalabilityWMRRDUIS}\includegraphics[scale=0.50] {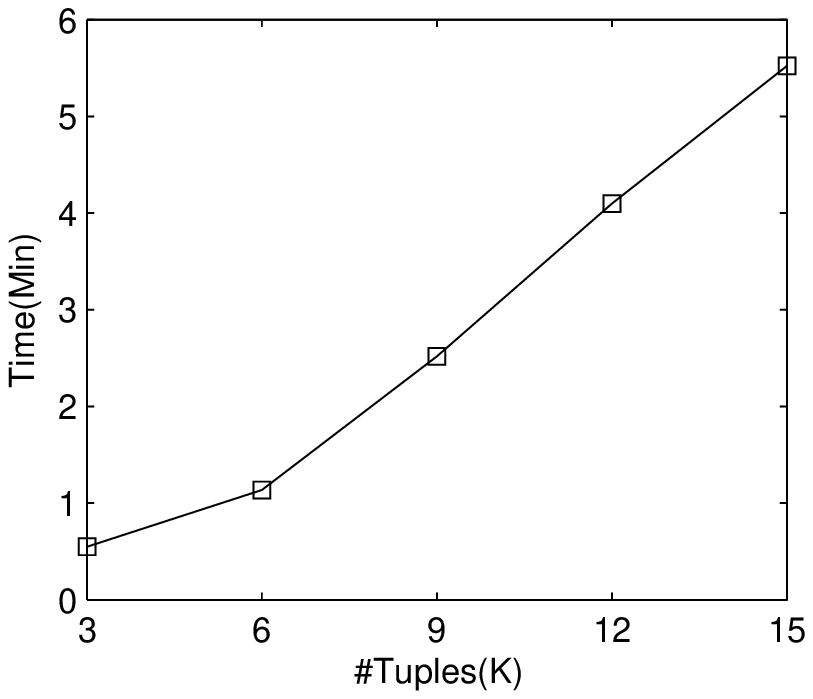}}
        \subfloat[Scalability of WMRRD on HOSP]{\label{figure:ScalabilityWMRRDHOSP}\includegraphics[scale=0.50] {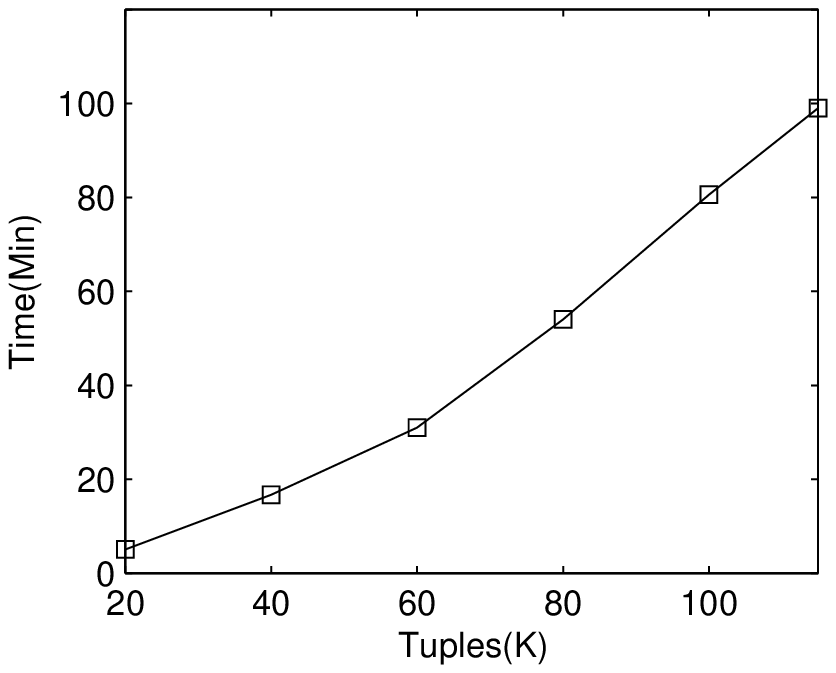}}\\
        \subfloat[Efficiency of Inconsis-Res on UIS]{\label{figure:EfficiencyInconsisResUIS}\includegraphics[scale=0.50] {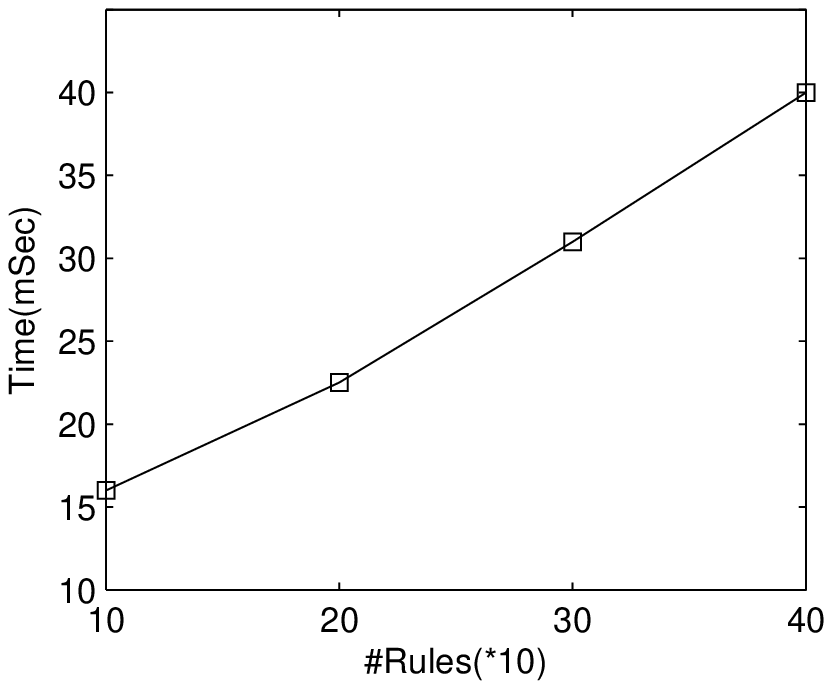}}
        \subfloat[Efficiency of Inconsis-Res on HOSP]{\label{figure:EfficiencyInconsisResHOSP}\includegraphics[scale=0.50] {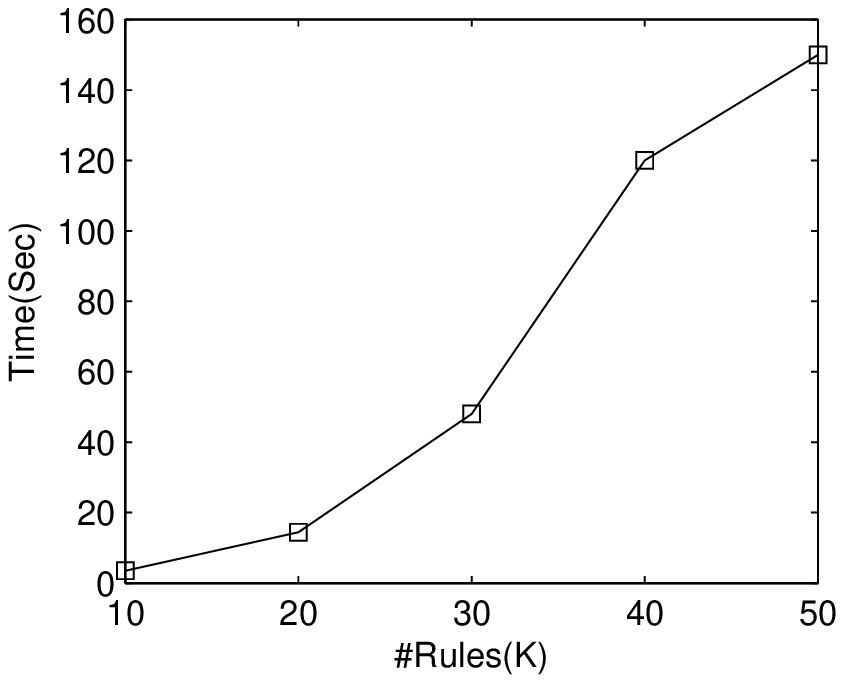}}\\
        \subfloat[Repairing Scalability on UIS]{\label{figure:RepairingScalabilityUIS}\includegraphics[scale=0.50] {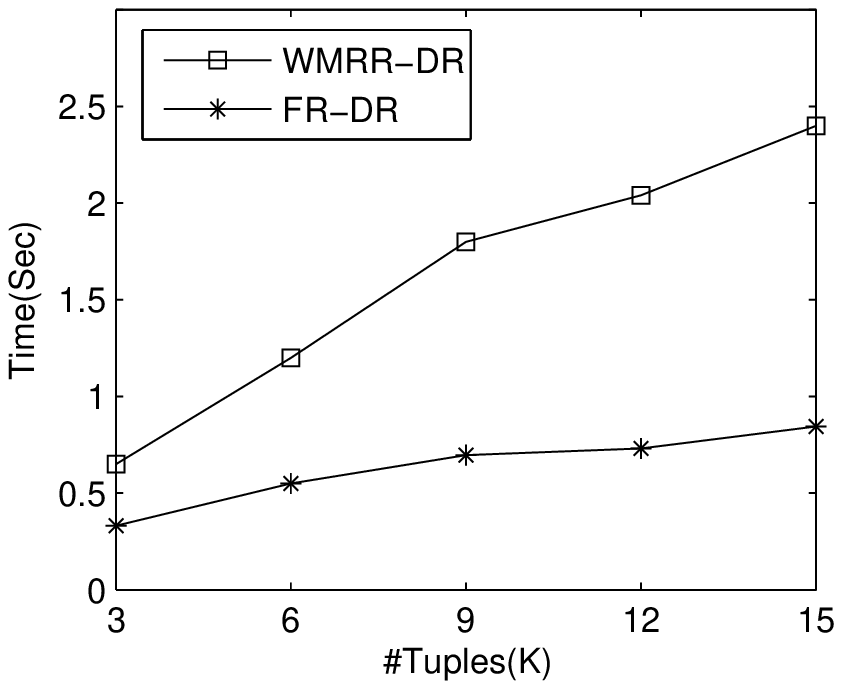}}
        \subfloat[Repairing Scalability on HOSP]{\label{figure:RepairingScalabilityHOSP}\includegraphics[scale=0.50] {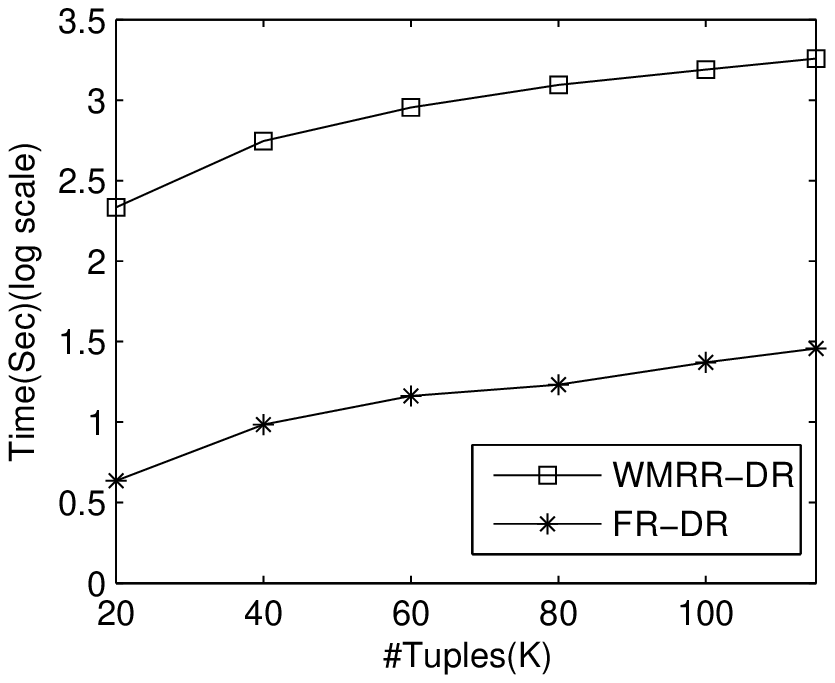}}
        \caption{Efficiency and Scalability}\label{figure:Efficiency and Scalability}
\end{figure}

Figs. \ref{figure:ScalabilityWMRRDUIS} and \ref{figure:ScalabilityWMRRDHOSP} show the runtime performance of WMRRD on UIS and HOSP, respectively.
They report that the runtime of WMRRD is approximately linear to the number of tuples on both data sets. This result shows that although the time complexity of WMRRD in the worst case is in quadratic in number of tuples (Sect. \ref{section:Rule Discovery}), it scales practically quite well.

Figs \ref{figure:EfficiencyInconsisResUIS} and \ref{figure:EfficiencyInconsisResHOSP} shows the runtime performance of Inconsis-Res on UIS and HOSP, respectively. 
The runtime of inconsistency resolution increases linearly on UIS with a small rule set size, and non-linearly on HOSP with a large rule set; where each pair of rules should be checked including all their wrong patterns. This non-linear result is not surprising because of the large rule set and the large number of negative patterns of rules that should be tested, where attribute values are highly frequent in HOSP.  However, Inconsis-Res scales well since it takes only 2.5 m to check and resolve inconsistency automatically in more than 58K rules on HOSP.

Figs \ref{figure:RepairingScalabilityUIS} and \ref{figure:RepairingScalabilityHOSP} depict a comparison between the run time of WMRR-DR and FR-DR on UIS and HOSP, respectively. It is not surprising that the run time of WMRR-DR with a large set of rules are higher than FR-DR with a small set of rules. Note that, there is a tradeoff between the accuracy and efficiency of WMRR-DR and FR-DR. As shown in figure \ref{figure:RepairingScalabilityUIS} and \ref{figure:RepairingScalabilityHOSP}, the repairing time of FR-DR is significantly lower than WMRR-DR; on the other hand, as illustrated in Table \ref{table: Avg-f-measure Comparison}, the repairing accuracy of WMRR-DR is significantly higher than FR-DR. Consequently, users can either repair a little number of data errors based on FRs with little time cost or repair a large number of data errors based on WMRRs with more time cost. As a result, our proposed method is preferred for 
real critical applications that care about high-quality data more than time cost, such that they can sacrifice some of time in order to save many costs caused by errors.

\emph{Summary.} From the experimental results, (1) our method achieves higher accuracy, compared with FR-DR, since it can achieve higher recall by rectifying more errors without any loss of repairing precision; (2) more rules are discovered by decreasing the threshold $\theta$ and, hence, the repairing accuracy is improved; (3) our method is more accurate than FR-DR even by using a little number of discovered rules. (4) WMRRD scales linearly with the size of data, Inconsis-Res scales well with the rule set size while there is a tradeoff between the accuracy and efficiency of WMRR-DR and FR-DR.

\section{Conclusion}
\label{section:conclusion}
In this paper, we introduce a new class of data repairing rules, weighted matching rectifying rules on which we can perform reliable data repairing automatically, based on the data in-hand without external data source or user verifications. We propose three effective algorithms to discover, check and apply these rules: (1) the rule discovery algorithm WMRRD which is the first algorithm to discover repairing rules automatically from dirty data in-hand, (2) the inconsistency resolution algorithm Inconsis-Res that checks rules consistency and also solve the captured inconsistency automatically, (3) the data repairing algorithm WMRR-DR that rectifies data errors based on the discovered rules. Our method is reliable, automatic, and highly accurate since it can rectify a large number of data errors correctly without user interaction or external data sources. We have conducted extensive experiments on both real-life and synthetic data sets, and the results demonstrate that WMRR-DR can achieve both high precision and high recall. This research is the first attempt to discover repairing rules automatically from the data in-hand utilizing correct values to repair errors without any external source. In future work, we would like to investigate techniques to reduce the number of discovered rules and enhance the repairing efficiency without loss of high-quality repairing.

\section*{Acknowledgements}
This paper was partially supported by NSFC grant U1509216, The National Key Research and Development Program of China 2016YFB1000703, NSFC grant 61472099,61602129, National Sci-Tech Support Plan 2015BAH10F01, the Scientific Research Foundation for the Returned Overseas Chinese Scholars of Heilongjiang Provience LC2016026. The authors would like to thank Prof. Jiannan Wang for their support in this work. 


\bibliography{References}
\end{document}